\documentclass{WileyMSP-template}

\usepackage{amsmath,amsfonts}
\usepackage[caption=false,font=footnotesize,labelfont=sf,textfont=sf]{subfig}
\usepackage{textcomp}
\usepackage{adjustbox}
\usepackage{pdfpages}

\usepackage{pgfplots}
\DeclareUnicodeCharacter{2212}{−}
\usepgfplotslibrary{groupplots,dateplot}
\usetikzlibrary{patterns,shapes.arrows}
\pgfplotsset{compat=newest}
\usepackage{tikz}
\usetikzlibrary{shapes,positioning,fit,calc}
\usepackage{circuitikz}
\usetikzlibrary{decorations}
\usetikzlibrary{decorations.pathmorphing}
\usetikzlibrary{decorations.pathreplacing}
\usetikzlibrary{decorations.shapes}
\usetikzlibrary{decorations.text}
\usetikzlibrary{decorations.markings}
\usetikzlibrary{decorations.fractals}
\usetikzlibrary{decorations.footprints}
\usetikzlibrary{babel}
\tikzset{>=latex}
\usetikzlibrary{arrows,automata}
\usetikzlibrary{math}

\title{Finite Element Simulation of NMC Particle Fracture during Calendering: a Route to Optimize Electrode Microstructures}

\date{June 2025}

\begin{document}

\pagestyle{fancy}
\rhead{\includegraphics[width=2.5cm]{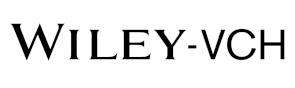}}

\maketitle

\author{Pierrick Guichard*}
\author{Benoit Mathieu} 
\author{Eric Woillez}


\begin{affiliations}
Dr. P. Guichard Univ. Grenoble Alpes, CEA, Liten, F-38000 Grenoble, France.\\
Email Address: pierrick.guichard@cea.fr

Dr. B. Mathieu Univ. Grenoble Alpes, CEA, Liten, F-38000 Grenoble, France.\\
Email Address: benoit.mathieu@cea.fr

Dr. E. Woillez Univ. Grenoble Alpes, CEA, Liten, F-38000 Grenoble, France.\\
Email Address: eric.woillez@cea.fr

\end{affiliations}



\begin{abstract}

Beyond active material intrinsic properties, the electrode manufacturing process is a crucial step to reach high energy density and long-life of Li-ion batteries. In particular, very high pressures are applied to the electrode during the calendering step, that directly influence the microstructure and the electrochemical performances. This article reports the first calendering simulation of a NMC cathode using a finite element method (FEM), including the post-fracturation behaviour of the secondary NMC particles. Calibrated with nano-indentation experiments, the mechanical model provides stress-strain predictions fully consistent with experimental data. On assemblies up to 100 particles, simulations reveal three calendering regimes along compression: particle rearrangement, moderate-pressure fracturing, and complete crushing. The model shows the strong sensitivity of the electrode microstructure to the calendering pressure level, and can thus be used as a guidance in the multi-criteria optimization of the manufacturing process.




\end{abstract}

\section{Introduction}

In the race for high energy density Li-ion batteries, NMC cathode materials have gained increasing attractivity thanks to their high potential ($\sim3,8V$) and high lithium storage capacity ($\sim275$ mAh/g) \cite{kim2018prospect,schmuch2018performance,yang2023ni}. Originally designed with a relatively high cobalt fraction (LiNi$_{1/3}$Mn$_{1/3}$Co$_{1/3}$O$_2$), manufacturers have progressively moved to Ni-rich materials (LiNi$_{0.8}$Mn$_{0.1}$Co$_{0.1}$O$_2$) to increase their energy density, and eventually get rid of cobalt, the latter being a critical and toxic material. The excellent electrochemical performances of Li-ion batteries with NMC cathodes has driven this technology to the leading market position for electric vehicles \cite{marchebatteries}. However, NMC cathode materials still have their drawbacks \cite{britala2023review}. Lower cobalt amount in the NMC results in a lower structural and thermal stability \cite{hua2019structural}, in particular due to oxygen release \cite{streich2017operando,jung2017oxygen,rinkel2022two}. The crystallographic structure is unstable when lithium is removed from the interstitial sites, during battery charge \cite{xu2021bulk}. The material then shrinks and can be irreversibly damaged if the lithium content drops below some threshold. Moreover, lithiation heterogeneities during battery cycling creates mechanical stresses that eventually lead to material fatigue and particle fracturation \cite{dose2021influence,hu2023challenges}.

To ensure fastest Li diffusion pathways and increased surface area, NMC materials are often synthesized with a hierarchical structure \cite{muller2021effect,trevisanello2021polycrystalline}. Nanosized homogeneous NMC crystals are gathered to form micro-sized secondary particles. Such structured NMC materials have revealed better capacity retention and longer lifetime than noncrystalline NMC. However, the material swelling creates inter-granular micro-cracks in which electrolyte can flow in and induce degradation reactions with the bare material surface. A careful adjustment of primary and secondary sizes is required to reduce the mechanical stress while keeping electronic percolation and material adhesion during long-term cycling \cite{wagner2020hierarchical}.

The battery performance does not only depend on the material synthesis, but it also strongly depends on the electrode manufacturing process that impacts its porous microstructure \cite{kremer2020manufacturing,NIRI2022100129}. After the electrode formulation has been chosen by mixing active and inactive materials, the manufacturing process is typically composed of a drying step where the solvant and possible residual water are removed. The electrode is then calendered at very high pressure - typically up to $200~\text{MPa}$ - to reach the desired porosity \cite{gunther2020classification,primo2021calendering}. The success of the manufacturing process has long relied on the know-how of cell makers, with manufacturing parameters adjusted using a trial-and-error procedure. Thanks to decades of experience, this approach has proved successful to build the very high quality Li-ion batteries that are now commercialized. It is however doubtful that the pure experimental approach can directly lead to the optimal manufacturing process, given the number of possible parameters and the complexity of the chemical and mechanical effects that result in the electrode microstructure. This is why numerical simulations, or so-called "digital twins", have attrackted more and more interest in the past few years to explore manufacturing routes that would not be easy or much too lengthy to realize with experiments \cite{ngandjong2021investigating,ayerbe2022digitalization}. One crucial issue is to rationalize the manufacturing process with unbiased and quantitative criteria.

The present work provides numerical simulations of the drying and calendering steps of the manufacturing process, with a special emphasis on particle fracturation. Most numerical approaches in the literature use the discrete element method (DEM) to reproduce microstructure mechanical properties \cite{stershic2015modeling,srivastava2020controlling,ayerbe2022digitalization,abdollahifar2023insights}, that does not take into account the intra-particle stress and strain distribution, nor does provide insights on the post-fracture behavior, as only rigid particles are considered. The DEM can be extended using clusters of primary particles linked with bonds to represent real-shape particles and to model particle fracture. Such methodology has recently been applied  to simulate the manufacturing process, including calendering, and predicts the resulting microstructure \cite{xu2023computational}. The drawback of such an approach is to loose the benefit of DEM in terms of computational time, and is thus restricted to a much smaller particle number. Moreover, the inevitable particle interface discretization with spheres raises questions about the correct contact representation. By contrast, the finite element method (FEM) can be used to simulate the internal stress and particle deformations by contacts, for particles with arbitrary shapes. Such approach has been recently initiated to model NMC fracturation during electrochemical cycling \cite{shishvan2023cracking}. A FEM approach is used in the present work to model the microstructure reorganization, the fracture initiation and particle cracking, with very few hypotheses on the fracturation model.  This allows a realistic simulation of the calendering process up to arbitrary high pressure levels.

The model is first applied to the fracture of a single secondary particle, validating the model in the linear regime and comparing the fracture properties of the material with literature results of nano-indentation experiments \cite{stallard2022mechanical} \cite{xu2019heterogeneous} \cite{wheatcroft2023fracture}.
Subsequently, the drying and calendering of a microstructure volume composed of 100 NMC secondary particles is simulated.
While the breaking point of NMC particles is above the calendering pressure, we show that the stress is not homogeneously distributed in the microstructure. NMC particles arrange themselves into force chains, also known as force networks, to withstand the pressure \cite{imseeh20183d}. This arrangement results in significant heterogeneities: some particles experience forces well below the average, while others are subjected to forces typically six times greater than the average. This explains why a large number of fractured particles  can already be observed right after the manufacturing process, which partly cancels the benefit of hierarchical NMC particles structure. The fracturation can lead to NMC island isolated from the electronic percolating network. It also increases the active surface in contact with the electrolyte and thus the surface layer reconstruction by oxygen release, that is known to be an important degradation mechanism for NMC materials.  

 The effective stress-strain relation of the microstructure is obtained up to fully damaged microstructures where all secondary particles have been decomposed in primary units. We present a qualitative and quantitative study of the evolution of fractured particles as a function of applied pressure, as well as key microstructure geometrical properties such as porosity and exchange surface. These results can be used to anticipate real microstructure properties and set manufacturing criteria to optimize NMC cathodes.

\section{Results and discussion}

Based on the microscopic properties of a secondary NMC particle, which consists of the aggregation of primary particles with a typical size of $\ell=0.5 ~\text{µm}$ \cite{anansuksawat2024uniform}, we propose modeling its fracture mechanics using a FEM model previously applied to granular rock fractures \cite{wu2018investigation}. After meshing the sphere representing a secondary NMC particle, cohesive elements are inserted at the interfaces of each volumetric element, mimicking the primary NMC crystals. These cohesive elements act like double-sided tape: they break definitively when a predefined fracture criterion is met, resulting in crack formation along the primary particle boundaries. The simulation is done with the software Abaqus. More information about the fracture model and the numerical simulation is provided in section \ref{section:methods}.

During the initial calibration phase, the validity of this fracture model is assessed by comparing it with experimental nanoindentation data obtained from single particles. Subsequently, a comprehensive workflow for microstructure generation is introduced and applied to qualitatively and quantitatively investigate the mechanics during the calendering phase. It suggests the presence of three regimes whose properties drive the mechanics of cathodes.

\subsection{Fracture of a single secondary particle and model calibration}

The fracture of NMC secondary particles is predominantly intergranular, occurring at the boundaries between primary crystals, and is thus governed by the adhesion properties of these elementary grains \cite{stallard2022mechanical}. For FEM simulations, these properties are induced by the parameters of the cohesive elements.
A common strategy for tuning the zero-thickness cohesive elements involves a trial-and-error approach, where parameters are adjusted until the macroscopic behavior matches experimental data.

 \begin{figure}[ht!]
    \centering
     \subfloat[Flat platen indentation.]{\begin{adjustbox}{clip,trim=0cm 0.cm 0cm 0cm,max width=0.23\linewidth}\includegraphics{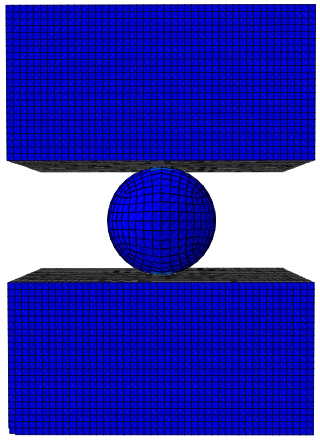}
\end{adjustbox}}\quad
\subfloat[Cono-spherical indentation.]{\begin{adjustbox}{clip,trim=0.cm 0.cm 0cm 0cm,max width=0.23\linewidth}\includegraphics{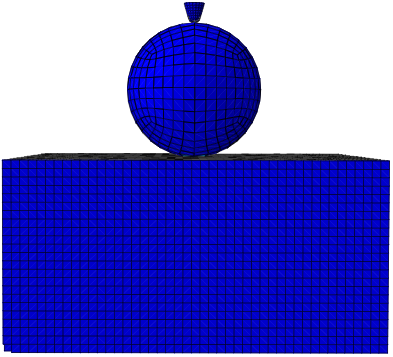}
\end{adjustbox}}\quad 
\subfloat[Particle fracture under flat platen indentation.]{\begin{adjustbox}{clip,trim=0cm 0.cm 0cm 0cm,max width=0.17\linewidth}\includegraphics{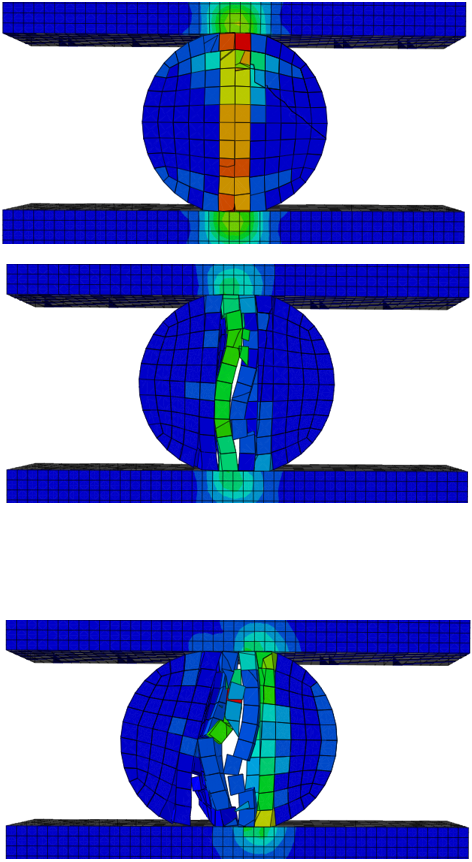}
\end{adjustbox}}\quad 
\subfloat[Particle fracture under cono-spherical indentation.]{\begin{adjustbox}{clip,trim=0cm 0.cm 0cm 0cm,max width=0.18\linewidth}\includegraphics{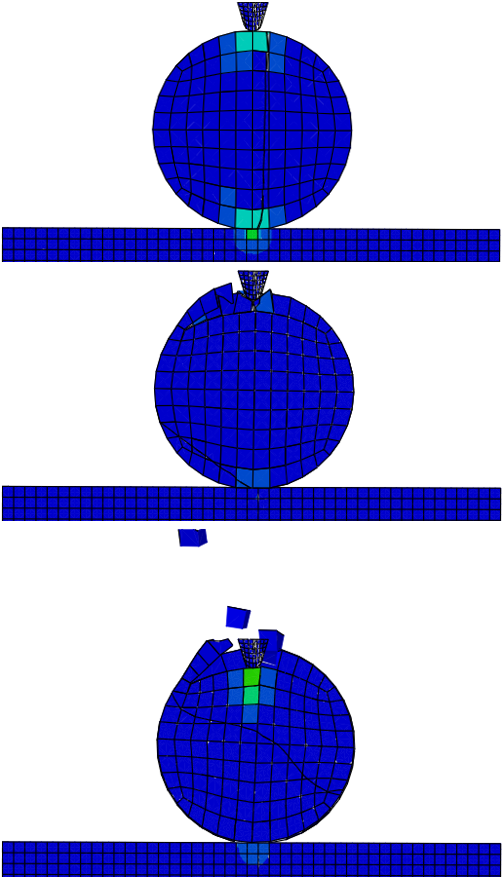}
\end{adjustbox}}
 \caption{\textbf{Numerical simulations of nanoindentation:} for the flat-platen indentation (a) as well as the cono-spherical indentation (b), the NMC particle is pressed by a diamond indenter on a silicium plate.
 The higher stress values are mostly located at the center of the particle (c-d), which is therefore the place where the cohesive elements get degradated the most, and where cracks open first, leading to the particle fracture.}
 \label{indent_devices}
\end{figure}

In this work, the experimental data originates from nanoindentation measurements of NMC secondary particles \cite{wheatcroft2023fracture}. They involve two different indentation devices, presented in figure \ref{indent_devices}: the flat platen indentation (a) and the cono-spherical indentation (b). For both devices, the bottom plate is made of silicon, with a Young modulus of $E_\mathrm{si} = 150~\text{GPa}$ and a Poisson ratio of $\nu_\mathrm{si} = 0.3$. The flat platen and cono-spherical indenters are both made of diamond, with a Young modulus of $E_\mathrm{diam} = 1141~\text{GPa}$ and a Poisson ratio of $\nu_\mathrm{diam} = 0.07$. The tip of the cono-spherical indenter has a radius of $R_\mathrm{ind} = 0.71$ µm. The data coming from this experimental work, as well as theoretical results valid in the linear regime, serves as a reference against numerical simulations for calibrating the values of the cohesive element parameters.

 \begin{figure}[ht!]
    \centering
      \subfloat{\begin{adjustbox}{clip,trim=0.cm 5.cm 0cm 0.cm,max width=1\linewidth}\includegraphics{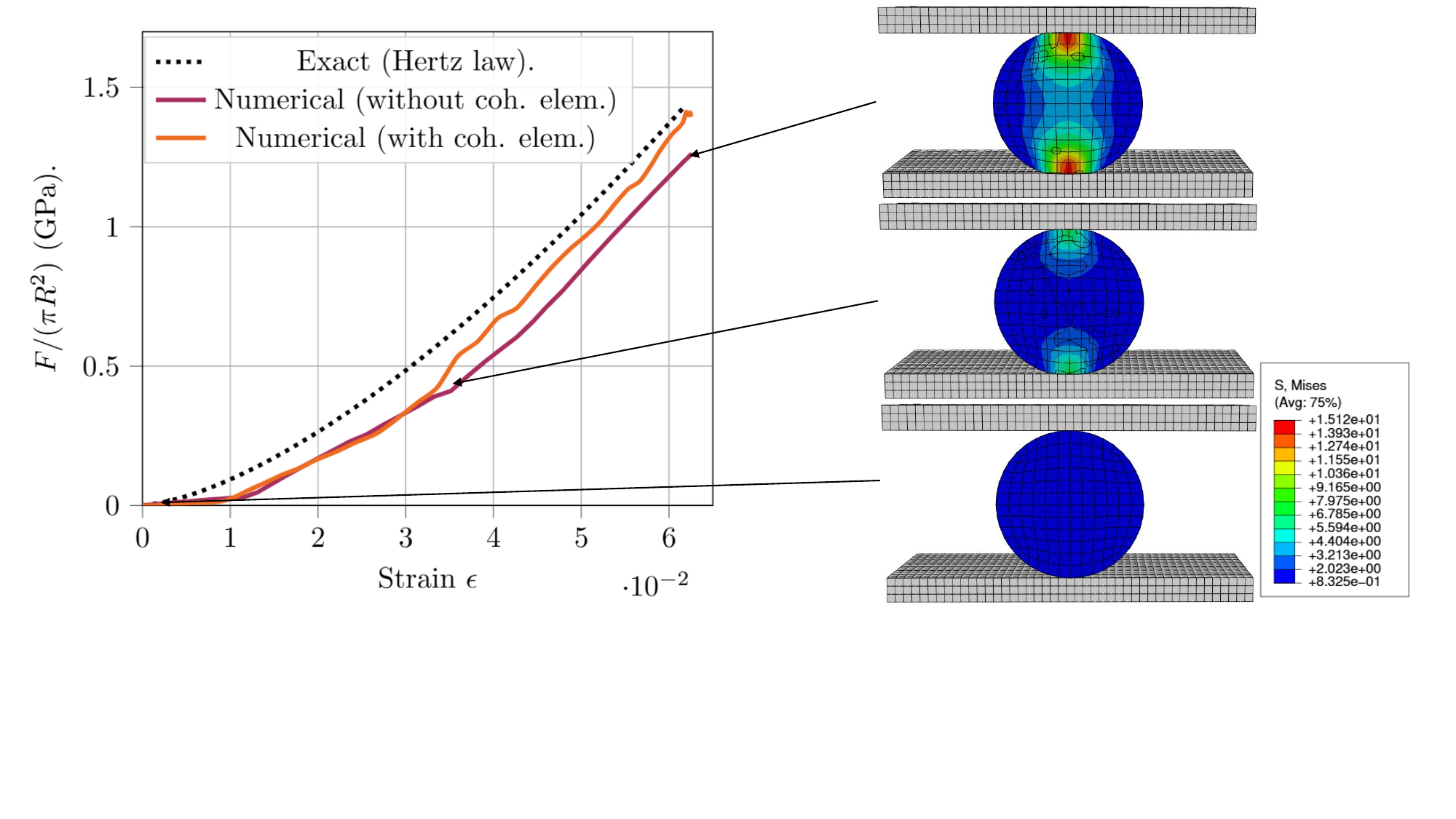}
 \end{adjustbox}}
 \caption{\textbf{Calibration of the fracture model parameter $k$ (the cohesive element stiffness) in accordance with the theoretical Hertz contact law.} On the left panel, the result given by the Hertz law is compared with the result coming from the sphere without cohesive element inserted (pure bulk material), and with the result coming from the NMC particle with non-degradated cohesive elements inserted (stiffness $k = 2500 ~\text{GPa.µm}^{-1}$ and $t^\mathrm{max} = +\infty$). On the panels on the right is represented the Von-Mises stress, during the compression of the NMC particle composed of pure bulk material.  }
 \label{fig:hertz}
\end{figure}


A cohesive element obeys a piecewise linear traction-separation law defining the degradation of the element, represented on figure \ref{coh_insert}. It gives the relationship between the components $t_n$, $t_s$ and $t_t$ of the traction vector ($t_n$ is the normal component to the cohesive element and $t_s$ and $t_t$ are the two tangential directions) and the corresponding displacements $\delta_n$, $\delta_s$ and $\delta_t$. Since we choose a isotropic model, we drop the indexes $n$, $s$ and $t$ in the rest of the paper. For any direction, the traction separation law, represented on figure \ref{coh_insert} is

\begin{equation}
    t = \begin{cases}
    &k \delta, \text{ if } \delta \leq \delta^\mathrm{init}\\
   & (1-D)k \delta, \text{ if } \delta^\mathrm{init}\leq \delta\leq \delta^\mathrm{fail}\\
   &0 \text{ otherwise. } \\
    \end{cases}.
    \label{coh_bilin}
\end{equation}

In this formula, $k$ is the stiffness of the cohesive element, and
$D := \frac{\delta^\mathrm{fail} (\delta - \delta^\mathrm{init})}{\delta(\delta^\mathrm{fail} - \delta^\mathrm{init})}$
is a scalar damage variable, increasing linearly from $0$ to $1$ with the displacement $\delta$ as the element fails. $D$ is a function of $\delta^\mathrm{init}$ the displacement at which the element starts failing, and $\delta^\mathrm{fail}$, the displacement obtained at the complete failure of the element. Equivalently, this constitutive law can be expressed in terms of the fracture energy $G$ (the amount of energy per unit area to open a crack), which is the integral of the bilinear curve defined by equation \ref{coh_bilin}, and of the value of the peak traction at $\delta^\mathrm{init}$, noted $t^\mathrm{max}$ (see figure \ref{coh_insert}). The advantage of this representation is that the value of $G =2~\text{J.m}^{-2}$ is readily given by the literature \cite{stallard2022mechanical} \cite{shishvan2023cracking}. Eventually, the cohesive element model is fully described by the parameters $k$, $t^\mathrm{max}$, $G$, and its density $\rho_\mathrm{coh}$, which are now to be adjusted.

A first criterion for the parametric calibration is that the cohesive elements should not artificially affect the properties of the bulk NMC material, when fracture is deactivated. In other words, the global macroscopic behavior of an NMC particle with inserted cohesive elements should match that of an NMC particle composed entirely of bulk material, when no fracture occurs in either case. Note that for preventing the cohesive elements from failing, $t^\mathrm{max}$ is temporarily set to infinity (in practice, a large value $t^\mathrm{max}\gg 1$). The contact law of a sphere of plain material is well known and serves here as a reference for adjusting the stiffness $k$. This theoretical result is called the Hertz-contact law \cite{johnson1982one}, and holds in the linear regime.
It exactly describes the force-displacement relationship of a particle pressed between two plates

\begin{equation}
    \frac{F}{\pi R^2} = \frac{4}{3\pi} \frac{E^*_1 E^*_2}{\left[(E^*_1)^{\frac{2}{3}} + (E^*_2)^{\frac{2}{3}}\right]^{\frac{3}{2}}}  \left( \frac{d}{R} \right)^\frac{3}{2}.
    \label{hertz}
\end{equation}

In this equation, $E_1^* := \left[\frac{1-\nu_{11}^2}{E_{11}} + \frac{1-\nu_{12}^2}{E_{12}}\right]^{-1} $ and $E_2^*:= \left[\frac{1-\nu_{21}^2}{E_{21}} + \frac{1-\nu_{22}^2}{E_{22}}\right]^{-1}$  are the equivalent Young modulus for the two contacts involving the Young modulus $E_{11}$, $E_{12}$, $E_{21}$, $E_{22}$ and Poisson ratios $\nu_{11}$, $\nu_{12}$, $\nu_{21}$ and $\nu_{22}$. $F$ is the force applied on the particle by the upper plate, and $d$ is the displacement of the upper plate.

When the two contacts on either side of the NMC particle are identical (i.e., $E_1^* = E_2^* := E^*$), equation \ref{hertz} simplifies to $\frac{F}{\pi R^2} = \frac{4}{3\pi} E^*  \left( \frac{d}{2R} \right)^\frac{3}{2}$. This simpler configuration is adopted for calibrating the stiffness $k$, as illustrated in figure \ref{fig:hertz}, which shows the compression of an NMC particle between two rigid plates. The relationship between pressure and strain is plotted on the left panel, while the right panels show cross-sections of the stress field. We compare the exact results predicted by the Hertz contact theory (black curve) with the results from a numerical simulation that does not include cohesive elements (purple curve). Subsequently, we set the stiffness of the cohesive elements so that after their incorporation into the numerical model, the pressure-strain curve (orange curve) matches the two previous curves (black and purple).
We find that the results are well approximated when the cohesive elements have a stiffness $k = 2500 ~\text{GPa.µm}^{-1}$ and peak load $t^\mathrm{max} = +\infty$. We also check that the stress field approximates well those obtained without cohesive elements plotted on the left panels of figure \ref{fig:hertz}. Therefore, this value of $k$ is adopted for the remainder of the paper.

Using the same approach, the density $\rho_\mathrm{coh}$ of the cohesive elements is now adjusted. The criterion remains that the undamaged response of the NMC particle should be consistent whether or not cohesive elements are present. Therefore, $t^\mathrm{max}$ is kept equal to infinity for this step. In combination with the stiffness $k$, the cohesive density $\rho_\mathrm{coh}$ primarily influences the propagation velocity of stress waves within the active material. Numerical simulations show that $\rho_\mathrm{coh}  = 0.1~\text{g.cm}^{-3}$ is appropriate for maintaining the same velocity in the NMC particle with cohesive elements, when compared to a plain NMC particle composed of pure bulk material.

 \begin{figure}[ht!]
    \centering
         \subfloat{\begin{adjustbox}{clip,trim=0.cm 0.cm 0cm 0cm,max width=1\linewidth}\includegraphics{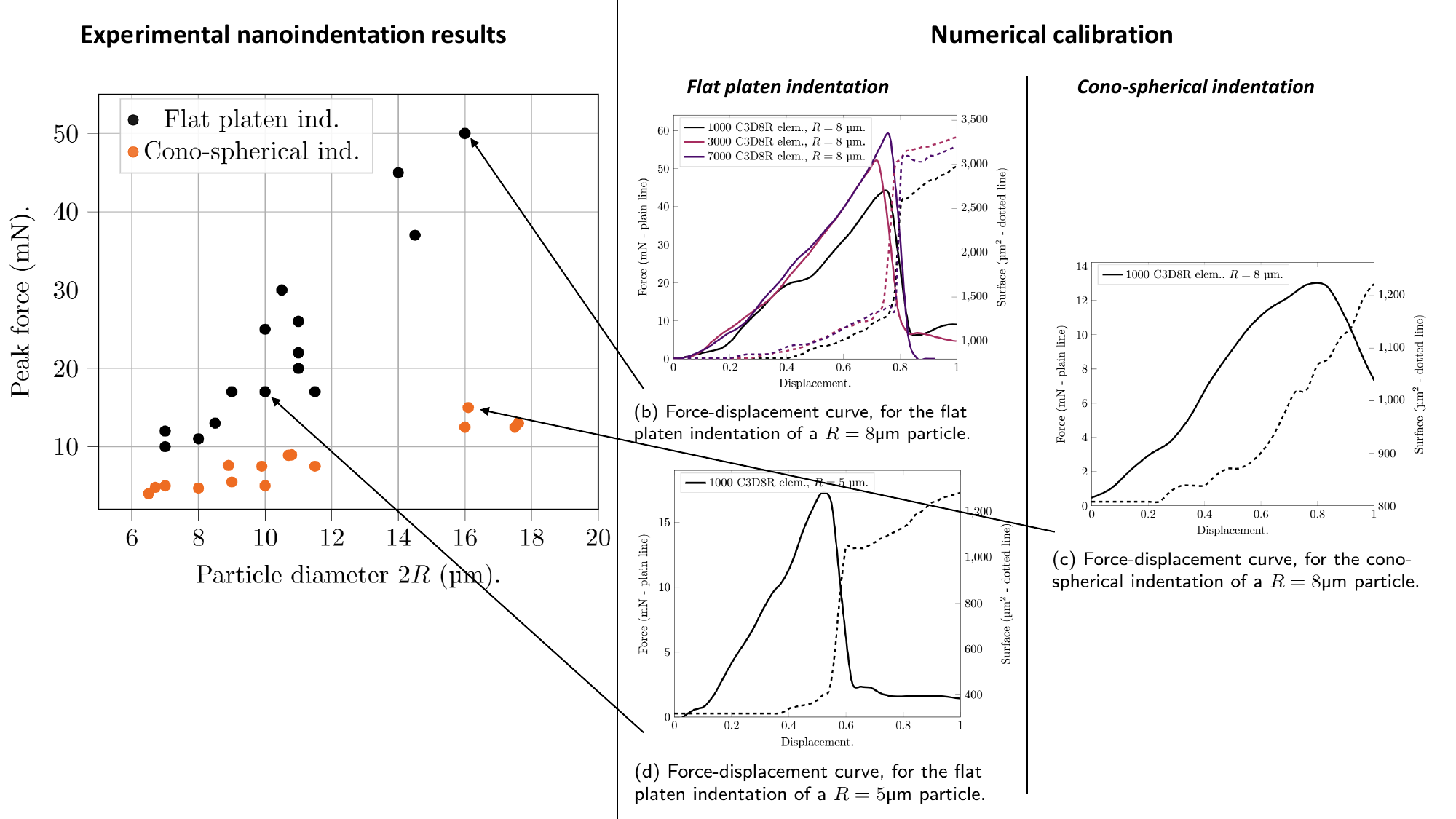}
\end{adjustbox}}\\
 \caption{\textbf{Calibration of the fracture model parameter $t^\mathrm{max}$ with respect to experimental nanoindendation measurements.} Left panel: experimental results for the peak force (force at fracture) in function of particle diameter, for the two indentation devices \cite{wheatcroft2023fracture}. Right panels: force-displacement curves, after adjustment of the cohesive element parameters. For these settings (which are the same on panels (b), (c), and (d)), the peak force matches the experimental data.}
 \label{nanoindent}
\end{figure}

A second criterion is that cohesive elements should fail for a correct amount of force, so that the numerical nanoindentation curves match the experimental data. In our fracture model, the failure initiation is controlled by $t^\mathrm{max}$, which is now adjusted in accordance with experimental results. The left panel of figure \ref{nanoindent} shows experimental results for the value of the peak force (the force at fracture), as function of the particle diameter. The black dots correspond to flat-platen indentations, while the orange dots are obtained with a cono-spherical indenter. A parabolic regression of these two curves sets a target for the fracture strength $\frac{F}{(2R)^2}$ around $0.2~\text{GPa}$ for the flat-platen indentation and $0.06~\text{GPa}$ for the cono-spherical device \cite{wheatcroft2023fracture}. On the right panels are presented the numerical calibration curves for NMC particles of different sizes, for both flat-platen and cono-spherical indentation. The plain curves represent the force as function of displacement, while the dotted curves show the surface as a function of displacement. For the $R=8~\text{µm}$ particle, the effect of the mesh density is also investigated: the black curve is obtained with $1000$ C3D8R elements, while the purple and dark purple curves show the results for $3000$ and $7000$ elements. All the curves were obtained with  $t^\mathrm{max}=0.4 ~\text{GPa}$, the value reproducing the experimental results of the left panel the most accurately.

Tests show indeed that a value of $t^\mathrm{max} = 0.4 ~\text{GPa}$ gives correct macroscopic fracture properties for the compression of a $8$ µm NMC particle under flat-platen indentation, the peak force being $F\simeq 50 ~\text{mN}$, as predicted by the parabolic regression of the experimental results. With the same value of $t^\mathrm{max}$, we check that the results are also correct for the cono-spherical indentation and for the indentation of a smaller $5~\text{µm}$ particle. Figure \ref{nanoindent} also shows that the exchange surface is multiplied by a factor larger than two during fracture. This gives a reliable criterion for detecting particle fracture with the sole knowledge of its exchange surface. Meanwhile, mesh dependency appears to have a lesser impact on fracture properties. Since using $1000$ C3D8R elements instead of $7000$ reduces the computational time by a factor of between $5$ and $7$, the coarser mesh is retained for the subsequent sections of this article.

\begin{table}[h!]
\centering
\begin{tabular}{c c c c} 
 \hline
 Parameters & Symbol (unit) & Values & Source \\ [0.5ex] 
 \hline
 NMC Young modulus & $E_\mathrm{NMC}$ (GPa) & 200 & \cite{stallard2022mechanical} \cite{sharma2023nanoindentation} \\ 
 NMC Poisson ratio & $\nu_\mathrm{NMC}$ & 0.3 & \cite{stallard2022mechanical} \cite{sharma2023nanoindentation} \\
 NMC density & $\rho_\mathrm{NMC}$ ($\text{g.cm}^{-3}$) & 4.8 & \cite{cheng2017mechanical} \\
 Secondary particle radius & R (µm) & $\simeq 6.5$ & \cite{stallard2022mechanical} \\
 Primary particle size & $\ell$ (µm) & $\simeq 0.5$ & \cite{anansuksawat2024uniform} \\
  Fracture energy & $G$ ($\text{J.m}^{-2}$) & 2 & \cite{stallard2022mechanical} \cite{shishvan2023cracking} \\
 Diamond indenter Young modulus & $E_\mathrm{diam}$ (GPa) & 1141 & \cite{stallard2022mechanical}\\
  Diamond indenter Poisson ratio & $\nu_\mathrm{diam}$  & 0.07 & \cite{stallard2022mechanical}\\
  Cono-spherical typ radius & $R_\mathrm{ind}$ (µm) & 0.071 & \cite{stallard2022mechanical}\\
   Silicon plate Young modulus & $E_\mathrm{si}$ (GPa) & 150 & \cite{stallard2022mechanical}\\
  Silicon plate Poisson ratio & $\nu_\mathrm{si}$  & 0.3 & \cite{stallard2022mechanical}\\
 Steel plate Young modulus & $E_\mathrm{steel} $ (GPa) & 200 & \cite{chen2016variation}\\
 Steel plate Poisson ratio & $\nu_\mathrm{steel}$ & 0.28 & \cite{chen2016variation}\\
 Steel plate density & $\rho_\mathrm{steel}$  ($\text{g.cm}^{-3}$) & 7.8 & \cite{chen2016variation}\\
  Alu. plate Young modulus & $E_\mathrm{alu} $ (GPa) & 68 & \cite{gong2014determining}\\
 Alu. plate Poisson ratio & $\nu_\mathrm{alu}$ & 0.33 & \cite{gong2014determining}\\
 Alu. plate density & $\rho_\mathrm{alu}$  ($\text{g.cm}^{-3}$) & 2.7 & \cite{gong2014determining}\\
 Assembly compression speed & $v_\mathrm{plate}$ ($ \text{µm.s}^{-1}$) & $ 2.10^{7}$ & -\\
 Friction coefficient & $\mu$ & 0.9-0.3 (Sedimentation-calendering) & \cite{sun2024elucidating} \\
 Cohesive element stiffness & $k$ ($\text{GPa.µm}^{-1}$) & 2500 & Trial and error.\\
 Cohesive element density & $\rho_\mathrm{coh} $ ($\text{g.cm}^{-3}$) & 0.1 & Trial and error.\\
 Cohesive element peak traction & $t^\mathrm{max}$ (GPa) & 0.4  & Trial and error.\\
  [1ex] 
 \hline
\end{tabular}
\caption{\textbf{Model parameters}.}
\label{table:1}
\end{table}

 A visual example of a particle fracture during the nanoindentation process is presented in figure \ref{indent_devices} (c-d). It results from simulations involving the parameters calibrated in the previous paragraphs (listed in table \ref{table:1}), and shows a cross-section of the particle before, during, and after fracture. For the flat platen indentation, before the fracture, stress is mostly localised in the center of the particle, on the z-axis joining the two contact areas. In this zone, the cohesive elements have already failed, but enough cohesive elements remain intact at a longer distance from the z-axis to ensure the macroscopic cohesion of the NMC particle. Then, the amount of failed elements increases, meaning that the particle starts opening from the center. As a result, it splits between several parts as the compression process continues. We observe, in accordance with experimental results, that the flat-platen and the cono-spherical indentation are relatively different. While the macroscopic fracture of the particle is achieved for lower levels of compression of the cono-spherical indenter, the fracture is less sudden than with the flat-platen. This is due to the sharpness of the indenter, which creates shear stress in the particle, and early microscopic cracks, but lacks an important contact zone for provoking the simultaneous failure of a large number of cohesive elements.

\subsection{Workflow for particles sedimentation}

\begin{figure}[ht!]
    \centering
     \subfloat{\begin{adjustbox}{clip,trim=0.cm 5.cm 2cm 0.cm,max width=0.9\linewidth}\includegraphics{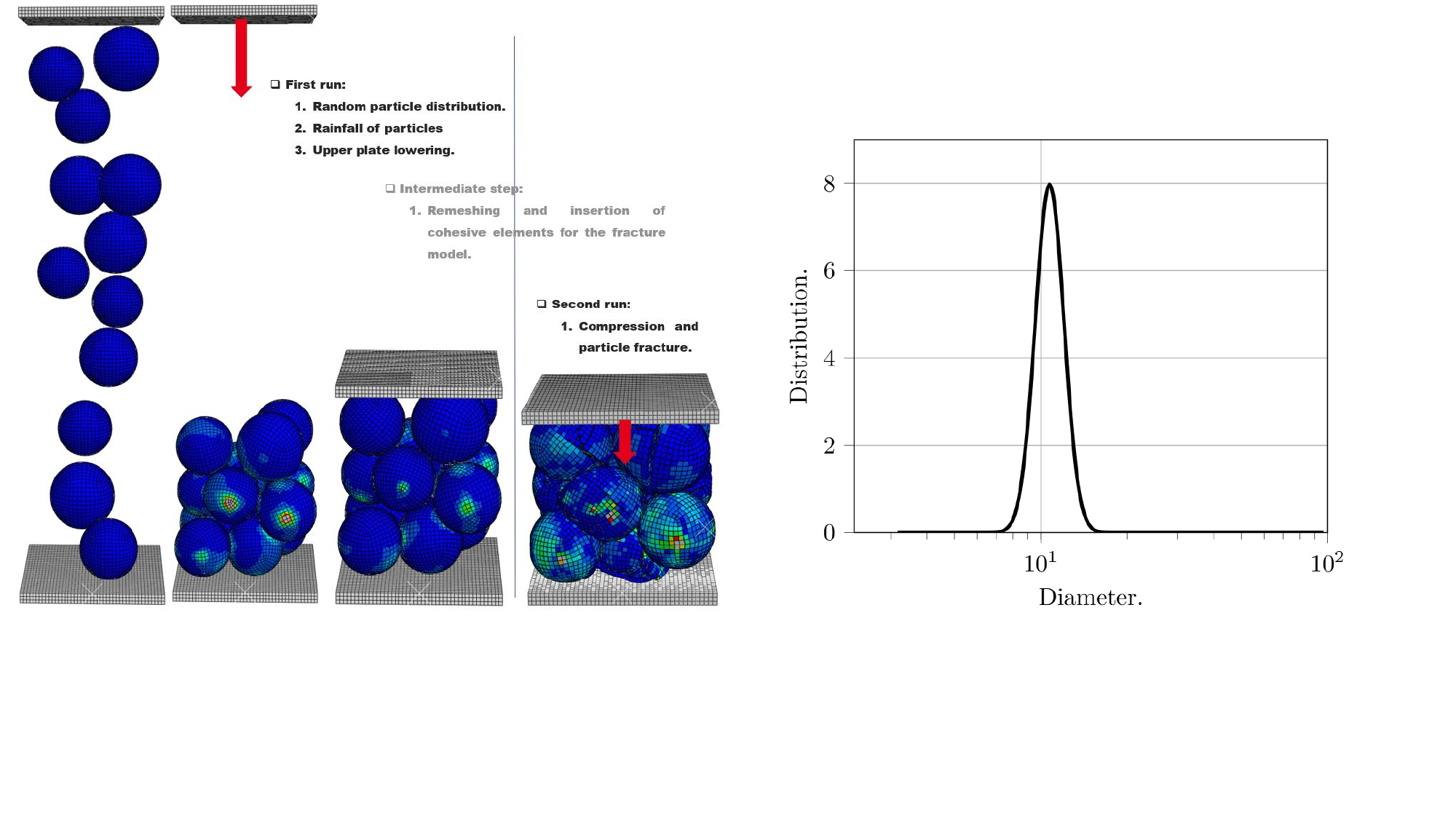}
\end{adjustbox}}
 \caption{\textbf{Description of the workflow for microstructure generation with Abaqus.} Left panel: a first inexpensive run ensures contact between the particles and the two compression plates. Then, the cohesive elements are inserted in the NMC particles. Eventually, compression is applied, leading to the fracture of some NMC particles. Note that here, the external mould maintaining the particles was removed for visualisation purposes. Right panel: distribution of the particle diameter.}
 \label{microstructure}
\end{figure}

This reliable fracture model of a single NMC particle, consistent with nanoindentation measurements is now integrated into a workflow of microstructure generation. This step is essential for investigating the fracture statistics inside a large volume representative of cathode active material.
Figure \ref{microstructure} represents the different steps of the microstructure generation, along with the particle size distribution. The whole process is separated in two distinct Abaqus runs.

\begin{enumerate}
    \item The first run gives a sedimentated microstructure of NMC particles, without cohesive elements. Initially, the particles of random size are randomly distributed in the volume between the two compression plate. Then, a gravity load is applied, resulting in a particle rainfall. A strong friction coefficient in the contact condition (typically, $\mu=0.9$) is employed so that mechanical energy is rapidly dissipated and the particles stop bouncing. Finally, the zero-displacement boundary condition is removed from the upper plate, which also falls onto the bed of particles. Note that this simulation is relatively unexpensive, since the number of elements is low. Also, non-physical parameters can be employed for accelerating the simulation, like softer and denser particles, a lighter upper plate, etc, since only the mesh properties are retained for the next Abaqus run. 
    \item At the end of the first phase, the next step is to automatically insert the cohesive elements into the NMC particles using the procedure described in the previous paragraphs and in section \ref{section:methods}.
    \item The second run consists in the compression of the active material bed, modeling the calendering procedure. It is fully described in section \ref{section:simu}.
\end{enumerate}


Note that separating the simulation in two distinct runs is necessary for decreasing the computational cost. Indeed, if the cohesive elements were inserted before the first run, the sedimentation phase would be very expensive, since it would involve around three times more elements with complex constitutive laws. It also allows running multiple calendering simulations for covering a wide range of parameters, while starting from the same microstructure configuration.


\subsection{Simulation of microstructure calendering}

\label{section:simu}

\begin{figure}[ht!]
    \centering
     \subfloat[First force chain.]{\begin{adjustbox}{clip,trim=0.3cm 0.cm 0cm 0cm,max width=0.287\linewidth}\includegraphics{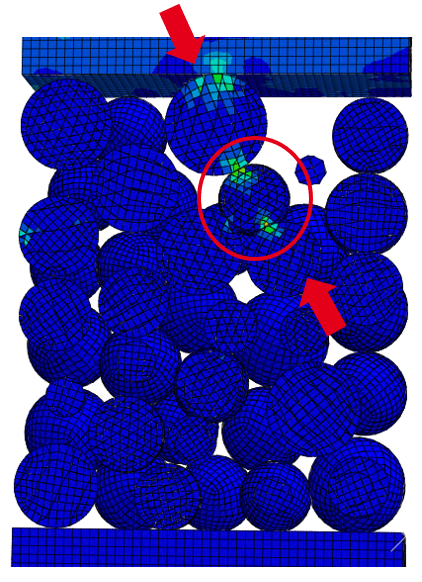}
\end{adjustbox}}\quad
\subfloat[Fracture.]{\begin{adjustbox}{clip,trim=0.3cm 0.cm 0cm 0cm,max width=0.295\linewidth}\includegraphics{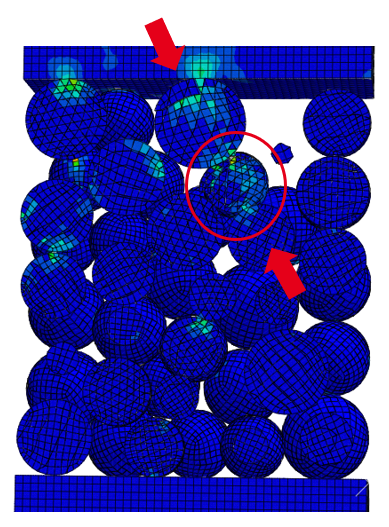}
\end{adjustbox}}\quad 
\subfloat[Second force chain.]{\begin{adjustbox}{clip,trim=0.3cm 0.cm 0cm 0cm,max width=0.3\linewidth}\includegraphics{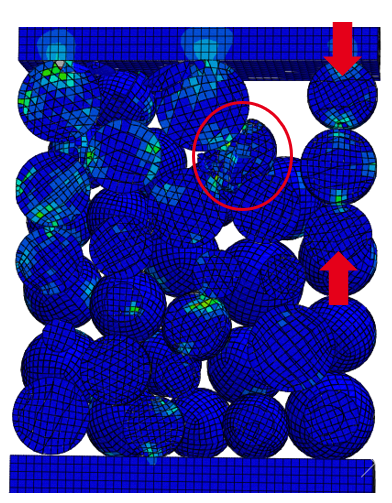}
\end{adjustbox}}
 \caption{\textbf{Reorganisation of the force chains} inside the NMC granular assembly. The high amount of force in the first force chain (left panel) leads to the fracture of the central particle (middle panel). Eventually, a new force chain (right panel) replaces the first one, due to the fracture of the particle in the red circle.}
  \label{force_chain_1}
\end{figure}

\begin{figure}[ht!]
    \centering
     \subfloat{\begin{adjustbox}{clip,trim=0cm 3.5cm 0cm 4cm,max width=1\linewidth}\includegraphics{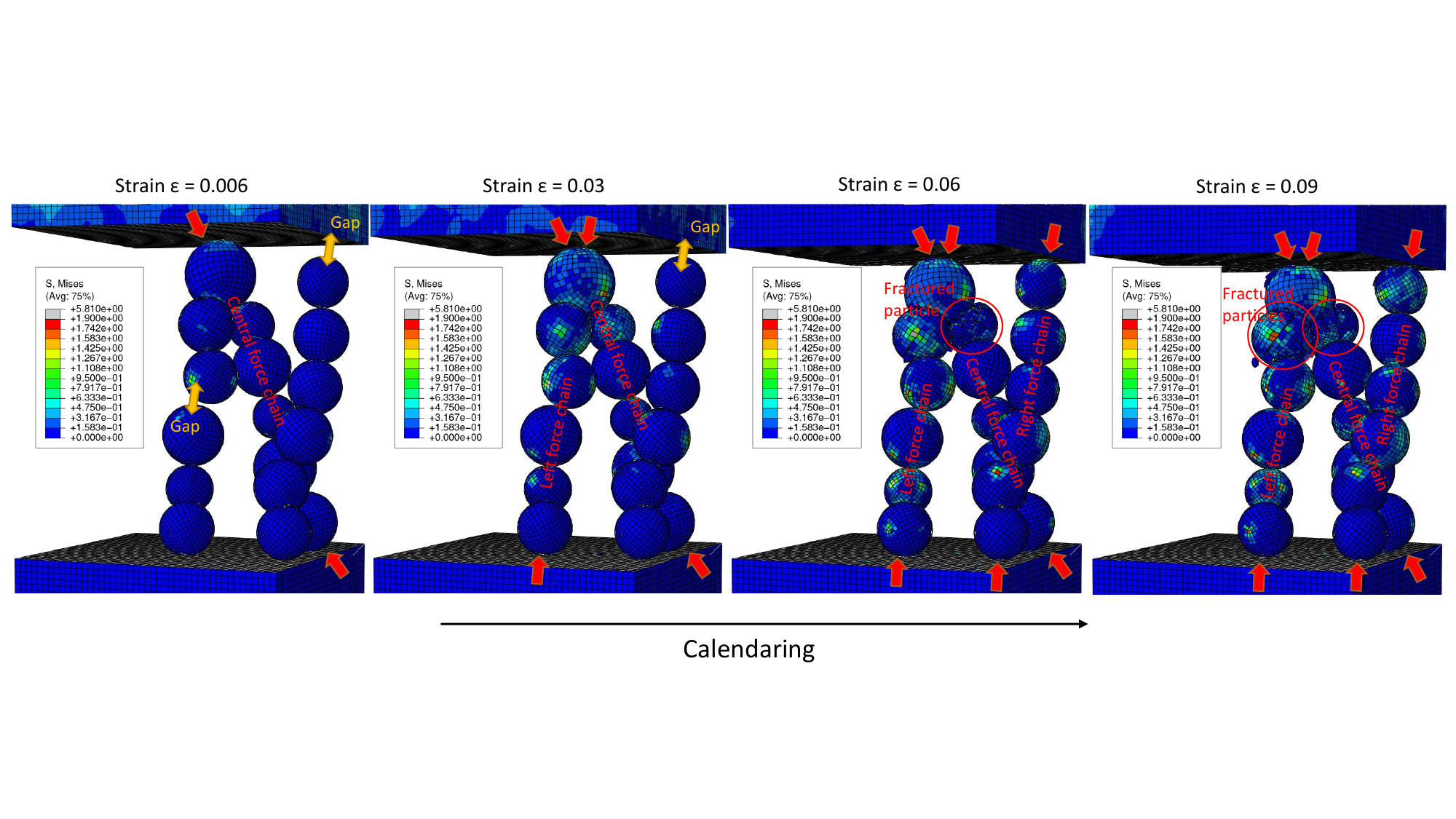}
\end{adjustbox}}
 \caption{\textbf{Identification and reorganisation of three force chains} in the granular assembly. Initially (for $\epsilon=0.006$), the upper compression plate is supported by the central force chain, the particles of the two other force chains are not in contact. Then, a particle of the central force chain begins deforming and fracturing (for $\epsilon=0.03$ and $\epsilon=0.06$): particles or the two extreme force chains are now in contact. Eventually (for $\epsilon=0.09$), the particles of the left force chain fracture as well: only the particles of the right force chain remains intact. }
  \label{force_chain_2}
\end{figure}

\begin{figure}[ht!]
    \centering
\subfloat{\begin{adjustbox}{clip,trim=7.cm 0.cm 8cm 0cm,max width=1\linewidth}\includegraphics{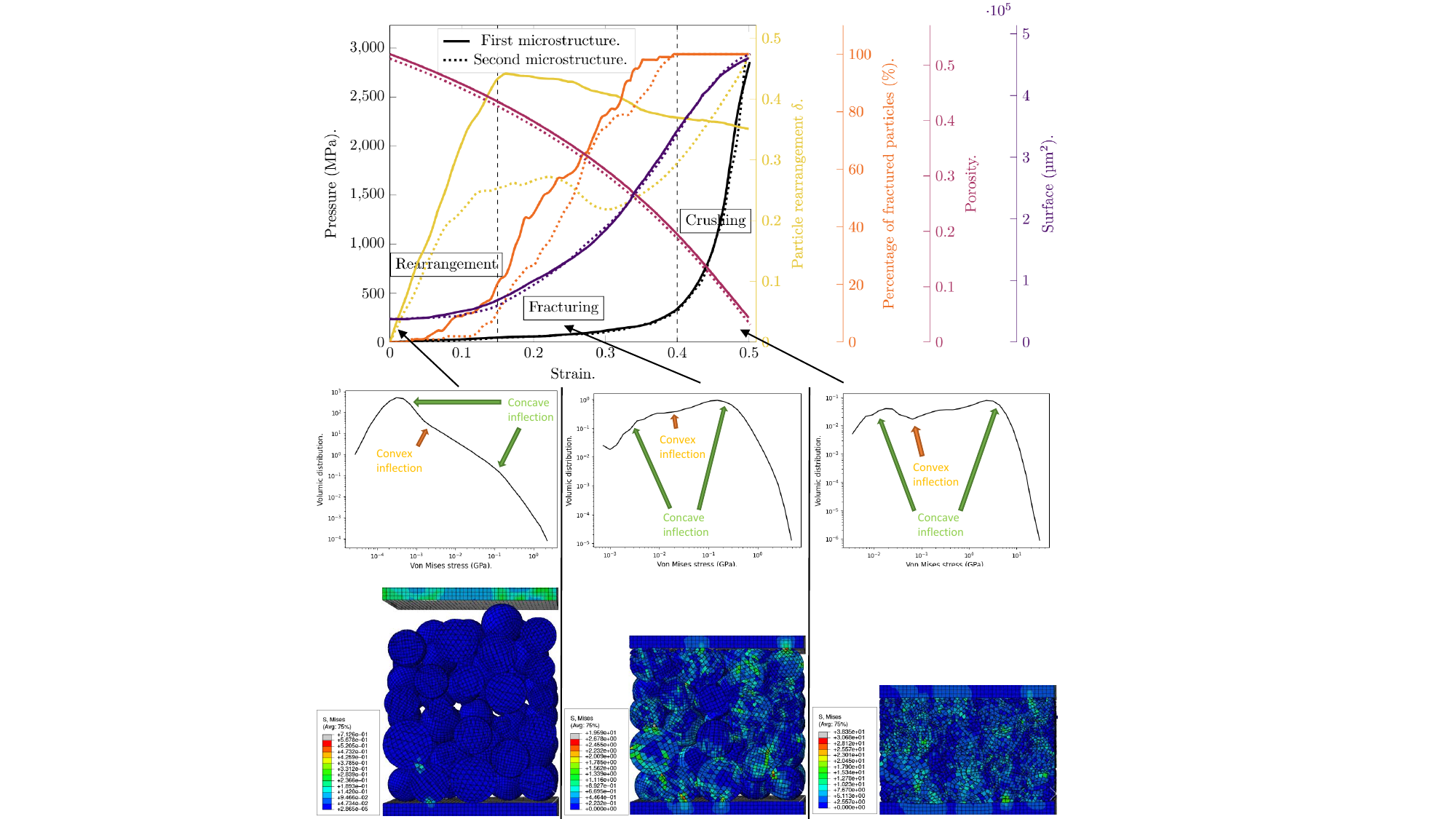}
\end{adjustbox}}
 \caption{\textbf{Calendering results for two different microstructures of 100 NMC particles.} The upper panel shows pressure, particle rearrangement, number of fractured particles, porosity, and surface in function of strain, for the two different randomly-sedimentated microstructures. The bottom panels display the distribution of Von Mises stress in the assembly for each mechanical regime using a log-log scale, along with a cross-sectional view of the sample.}
 \label{statistics}
\end{figure}

Applying the methods presented in the previous sections, we simulate the calendering of an assembly of 100 particles with average radius of $R=6.5$ µm and standard deviation of $2$ µm (the distribution is provided on figure \ref{microstructure}), and present the results on figures \ref{force_chain_1}, \ref{force_chain_2} and \ref{statistics}. The upper compression plate is made of steel with density $\rho_\mathrm{steel} = 7.8 ~\text{g.cm}^{-3}$, Young modulus $E_\mathrm{steel} = 200 ~\text{GPa}$ and Poisson ratio $\nu_\mathrm{steel} = 0.28$, while the bottom compression plate models the aluminium current collector of thickness $20$ µm, density $\rho_\mathrm{alu}=2.7 ~\text{g.cm}^{-3}$, Young modulus $E_\mathrm{alu} = 68 ~\text{GPa}$ and Poisson ratio $\nu_\mathrm{alu} = 0.33$. The compression speed of the upper plate is set to $v_\mathrm{plate} = 20~\text{m.s}^{-1}$, which is more than two orders of magnitude lower than the velocity of mechanical waves $\sqrt{\frac{E_\mathrm{NMC}}{\rho_\mathrm{NMC}}} \simeq 6455 ~\text{m.s}^{-1} $ in NMC particles. 
We additionally checked that $v_\mathrm{plate}$ is low enough to achieve a quasi-static compression, by comparing the reaction force on the upper and bottom plate, and ensuring that they do not present significant differences nor oscillations. Based on \cite{sun2024elucidating}, the friction coefficient was set to $\mu=0.3$ for the active material and compression plates. As only exception, we assumed a frictionless contact ($\mu  = 0$) between the external mould (device maintaining the particles together, not represented on the figures) and all the other parts of the assembly. Note that we performed calculations on two distinct assemblies of 100 NMC particles that were sedimented independently (plain and dotted curves of figure \ref{statistics}), to check the reproducibility of the results.

The thickness of the NMC layer is typically on the same order of magnitude as the diameter of ten individual particles. Consequently, the cathode active material exhibits significant microscopic heterogeneities due to its granular structure \cite{sun2010understanding, zhang2017role}. This results in the material not behaving like a continuous medium at the electrode scale. Specifically, the NMC particles organises themselves in force chains to resist the applied pressure, as presented on
figures \ref{force_chain_1} and \ref{force_chain_2}. These mechanical structures are composed of at least 3 linear connected particles presenting a level of stress well above average \cite{peters2005characterization}. Here, the chains predominantly align in a vertical direction and do not form a tree-like structure. This is because the pressure is applied along the z-axis and the granular layer is very thin. Between the force chains, other particles experience significantly lower stress levels. Typically, the particles within the force chains are referred to as the strong network, while the remaining particles are called the weak network. It has been demonstrated \cite{radjai1998bimodal} that the weak network contributes mostly isotropically to the stress tensor, effectively acting as a pressure. In contrast, the strong network is responsible for the deviatoric part of the stress tensor. In other words, the strong network serves as a skeleton that enables the granular assembly to resist shear stress and behave like a solid, while the weak network provides overall stability to this skeletal structure, preventing the force chains from collapsing laterally.

Usually, in the absence of fracturing, force chains are known to reorganize themselves during compression. Typically, the particles experience small relative displacements and rotations that redistributes the location of the chains. Here, this phenomenon is amplified by the fracture mechanism of the particles. When a high level of stress is encountered in a particle, it fractures. This relaxes the constraints in the neighbouring particles and therefore leads to a redistribution of the force chains. An example is provided in figure \ref{force_chain_1}. On the left panel, in the initial configuration, the upper plate is mostly supported by the force chain pointed by the arrows.  On the middle panel, the central particle of the chain fractures. As a result, the right panel shows that after the event, the dominant force chain is now pointed by the arrows at the right of the cross-section, while the particle of the previous force chains continues fracturing. Figure \ref{force_chain_2} presents a 3D view of the same phenomenon. Initially, the upper plate is supported by the central force chain. When one of its particles fractures, the two other chains are activated one after the other to withstand the pressure, before their particles also start fracturing.

The microscopic mechanical properties of the granular NMC assembly, as force chain rearrangement, directly impacts the properties of the sample at the system level, during the calendering process.
Typically, the manufacturer is interested in the values of the pressure, number of fractured particles, porosity, or exchange surface.
Figure \ref{statistics} shows their values for a  compression up to a strain of $0.5$. 
On this figure, the porosity $\frac{V_\mathrm{void}}{V_\mathrm{tot}}$ is defined as the void volume $V_\mathrm{void}$ divided by the total volume of the parallelepiped containing the assembly $V_\mathrm{tot}$. It is computed by adding together all the individual contributions of each volumic element. The exchange surface, defined as the contact area between the active material and the void, is calculated by summing the small contributions from each face in contact with the void. The upper panel also presents the evolution of the number of fractured particles. In accordance with the calibration curves of figure \ref{nanoindent}, a particle is declared fractured as soon as its total exchange surface has doubled.

The relationship between the microscopic mechanics and the system-level physics is provided by the rearrangement $\delta$, also plotted on figure \ref{statistics}. It estimates the particle relative movements during calendering and is defined by \begin{equation}
    \delta := \frac{1}{R}  <\sqrt{\left[x - x_0\right]^2 + \left[y - y_0\right]^2 +\left[z - z_0(1-\epsilon)\right]^2 }>.
    \label{rearrangement}
\end{equation}
In this formula, $R$ is the average radius of the particles ($6.5$ µm), $x$, $y$ and $z$ are the coordinates of the center of mass of each particle, $x_0$, $y_0$, $z_0$ are the initial coordinates of the center of mass of each particle, $<>$ is the average function over the complete set of NMC particles, and $\epsilon$ is the strain. Note that $\delta$ remains equal to zero throughout the simulation when every particle verifies $x = x_0$, $y=y_0$, and $z = z_0(1-\epsilon)$, therefore $\delta$ quantifies the ability of the particles to deviate from their locked position, performing micro-displacements to adapt to the applied pressure. 

In addition to these macroscopic scalar quantities, the real advantage of using a Finite Element software is its ability to compute the values of fields, like the stress field, almost continuously (up to the mesh discretization) within each particle. In fracture mechanics, the level of shear stress, responsible for particle cracking, is often evaluated through the Von Mises stress $\sigma_\mathrm{VM}$, a scalar value more practical to handle than the whole stress tensor $\sigma$. It is defined by
\begin{equation}
    \sigma_\mathrm{VM} = \sqrt{\frac{(\sigma_\mathrm{I} - \sigma_\mathrm{II})^2 + (\sigma_\mathrm{I} - \sigma_\mathrm{III})^2 + (\sigma_\mathrm{II} - \sigma_\mathrm{III})^2}{2}},
\end{equation}
where $\sigma_\mathrm{I}$, $\sigma_\mathrm{II}$ and $\sigma_\mathrm{III}$ are the principal stresses, in other words the eigenvalues of the stress tensor $\sigma$. Note that $\sigma_\mathrm{VM}$ is proportional to the second invariant $J_2$ of the deviatoric part of the stress tensor, i.e the part responsible for the distortion. Figure \ref{statistics} represents the volumic results of the Von Mises stress in the particle assembly for different levels of compression, in other words the histogram of the Von Mises stress of all the nodes. This result is given in a logarithmic scale for a better identification of the different regimes.

 \begin{figure}[ht!]
    \centering

\subfloat{\begin{adjustbox}{clip,trim=0.cm 0.cm 0cm 3cm,max width=1\linewidth}\includegraphics{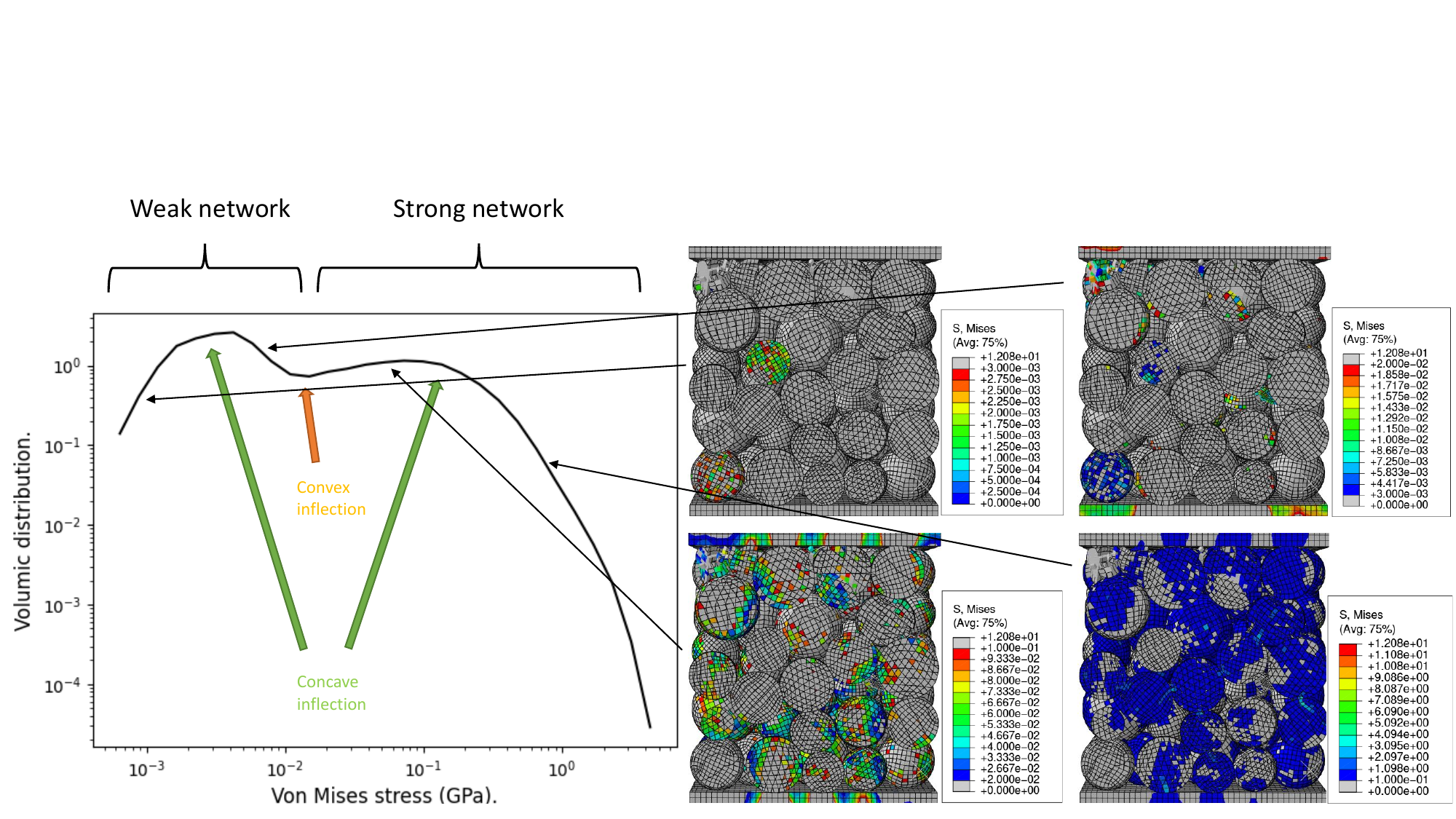}
\end{adjustbox}}
 \caption{\textbf{Histogram of the Von-Mises stress distribution and spatial localization of the stress levels.} On the left panel is plotted the volumic distribution of the Von Mises stress in the NMC particles, at strain $\epsilon = 0.17$. We note the presence of two local maxima (concave inflexion points) and one minimum (convex inflection point). The first peak corresponds to particles of the weak force network, while the second peak is mostly composed of particles of the strong network. This is confirmed by the cross-sections presented on the left of the figure. They show the localization of the stress levels, for each part of the histogram presented on the left panel.  }
 \label{hist_statistics}
\end{figure}

\subsection{Discussion}

The results presented in Figure \ref{statistics} illustrate the evolution of cathode properties during the calendering process for two distinct microstructures, differentiated by plain and dotted curves. We observe that the results for pressure, porosity, and exchange surface area are nearly identical for both microstructures. In contrast, the rearrangement indicator and the number of fractured particles show a higher level of variation between the two microstructures. This discrepancy arises because the former quantities (pressure, porosity, and exchange surface area) are well-averaged over the sample, whereas the latter indicators (rearrangement and particle fracture) are closely related to the microscopic properties of the structure. 

Overall, the mechanical properties of the sample differ significantly from those of individual NMC particles, a characteristic behavior of granular media. Notably, at a given strain, the pressure exerted is substantially higher for a single, intact NMC particle (as depicted in figure \ref{fig:hertz}) compared to a multiparticle assembly (figure \ref{statistics}). This discrepancy arises because fracturing and particle rearrangement within the assembly help to relax stress concentrations.
Another key difference is the pressure at which initial fracturing occurs. In a multiparticle assembly, the first particles fracture at approximately $50~\text{MPa}$, whereas single particles exhibit a peak pressure of around $200~\text{MPa}$ (figure \ref{nanoindent}). This difference can be attributed to the strong interparticle network within the assembly, which can locally concentrate stresses well above the average, exceeding the fracture threshold of individual particles.


On figure \ref{statistics}, all the variables suggest the presence of three regimes experienced by NMC particles during compression. The pressure curve shows that for strains lower that $0.4$, pressure remains relatively low (smaller than $250~\text{MPa}$), whereas it significantly increases for strains between $0.4$ and $0.5$. The rearrangement curve show that the low strain domain can be additionally separated in two distinct regions. For very low strains (lower than $0.15$), rearrangement increases linearly with strain, while is saturates for strains higher than $0.15$. This critical level of strain seems to match the strain required for fracturing a large number of particles. We also checked that particles rearrange mostly along the z-axis, during the early stages of compression, which corresponds to an evening of the upper surface. Indeed, after the drying step, the upper surface is often uneven. When pressure is applied, the upper particles sink into the granular bed. This sinking motion is compensated by the ascending movement of particles that are not yet in contact with the calendering roll.


The lower panels of Figure \ref{statistics} illustrate the Von Mises stress distribution during calendering, revealing both similarities and differences that aid in identifying distinct compression regimes. For each strain level, at high stress, the logarithm of the stress distribution $\mathcal{P}$ exhibits a linear relationship with the logarithm of the Von Mises stress $\sigma_\mathrm{VM}$, although this observation is not definitive. This relationship can be expressed as $\log(\mathcal{P}) = \alpha \log(\sigma_\mathrm{VM}) + \beta$, meaning that $\mathcal{P} \propto \sigma_\mathrm{VM}^\alpha$. The slope $\alpha$ decreases from approximately $-1$ to $-5$ during calendering, indicating that the distribution of high stresses becomes increasingly homogeneous across the sample as particles fracture.

Each stress distribution features two concave inflection points (which are sometimes local maxima), and one convex inflection point (which is sometimes a local minimum). Figure \ref{hist_statistics} provides a detailed identification of these curve segments at a strain level of $\epsilon=0.17$, chosen for the clarity of the inflection points, though the results are applicable to other strains. For each segment of the distribution, cross-sectional views of the sample precisely illustrate the localization of corresponding stress levels. To achieve this, only the stress levels relevant to each segment of the curve were selected and displayed according to the color scale on the cross section; stress levels exceeding the segment’s bounds are shown in gray.
Observations reveal that high stress levels (second peak) correspond to particles aligned vertically in a network structure, effectively resisting the applied pressure—this corresponds to the strong network of particles involved in force chains. Conversely, low stress levels (first peak) correspond to particles not directly involved in force chains but maintaining lateral stability of the chain structure, which corresponds to the weak network.
For both peaks, the increasing segment of the curve corresponds to nodes located at a certain distance from particle contacts and oriented in the normal direction of contact forces. In contrast, the decreasing segment relates to nodes in the immediate vicinity of particle contacts, aligned with the direction of contact forces.

These observations are crucial for analyzing the distributions in the lower panels of Figure \ref{statistics}. At lower compression levels, the first peak comprises a relatively high number of nodes, as the weak network primarily consists of complete secondary particles. In the second regime, the relative area of the first peak decreases, with the weak network composed of smaller fractured secondary particles, still larger than primary crystals. Indeed, the fracturing process relaxes stress in the corresponding particles, explaining why the lowest stress levels are observed in fragments rather than in complete particles. At the highest compression levels, the relative importance of the first peak diminishes further, consisting mainly of individual primary particles. This confirms that during calendering, NMC particles reorganize and fracture into smaller parts, eventually crushing into primary crystals. This process can be summarized in three regimes, as illustrated by figure \ref{statistics}.

\begin{enumerate}
    \item \textbf{Rearrangement:} low levels of strain (less than $0.15$). 
    Pressure is low (less than $50 ~\text{MPa}$). Particles perform micro adjustments and rotations to reorganise force chains, without large amounts of particle fracturing. Rearrangement primarily occurs along the z-axis, which means it mainly contributes to evening out the upper surface of the particle bed. 
    \item \textbf{Fracturing:} intermediate levels of strain (between $0.15$ and $0.4$). Pressure remains moderate (less than $150~\text{MPa}$). 
    The position of each particle is locked, but force chain redistribution still happens due to particle fracturing, keeping that level of stress moderate. Progressively, every particle fractures during this phase. 
    \item \textbf{Crushing:} important levels of strain (larger than $0.4$). 
    Pressure increases significantly, up to $3000~\text{MPa}$. The already fractured particles keep opening in smaller parts. Eventually, each volumic element is disconnected from its neighbours, transforming the initial secondary NMC particle assembly into a primary NMC particle assembly. 
\end{enumerate}

The macroscopic results presented in figure \ref{statistics}, in particular the pressure-strain curve, are relatively similar to other numerical models \cite{sun2024elucidating} that do not take particle fracture into account. A significant difference however is that the curve seems to be shifted towards the high strains. In other words, the important increase in pressure starts later for our model, which takes into account particle fracture. Fracture allows indeed a relaxation of the level of stress at a given level of strain.

In comparison to experimental results, our study reveals a higher incidence of particle fracture, with each particle failing at a strain close to $0.3$. This elevated fracture rate is likely due to the absence of a binder in our model, which might significantly influence the battery's mechanical behavior. Introducing a plastic binder into the model would likely alter particle interactions and distribute stress more evenly, thereby reducing the number of fractured particles. While incorporating a binder would enhance the model's realism, it would also necessitate substantial additional simulation efforts beyond the scope of this paper. The binder exhibits indeed complex viscoelastic behavior, and preliminary simulations indicate that its deformations, which differ considerably from those of the NMC material, pose challenges to mesh integrity.

It is also essential to remember that due to computational constrains, the size of the assembly remains small, thus the results have still to be extended to larger assemblies. In particular, more particles fracture near the analytical rigid boundary along the x and y axis, meaning that future simulations need to take this effect into account. Still, the three observed regimes were assessed on multiple simulations, for various particle distribution and seem to be consistent with experimental works, even if future research may refine the quantitative results.

This study in particular reveals that the typical calendering pressures selected by cell manufacturers, ranging from $100$ to $200~\text{MPa}$, fall within the mid-range of the fracturing regime depicted in Figure \ref{statistics}. This regime is characterized by the highest fracturation rate with respect to pressure increase, as shown by the percentage of fractured particles (orange curve) and the surface (blue curve). The simulations show that a reduction of the calendering pressure to 70 MPa could prevent the electrode to enter in the fracturing phase, which is desirable to mitigate subsequent electrochemical degradation phenomena. However, the pressure reduction also means a much higher porosity above 0.4, that do not meet the energy KPIs of standard battery electrodes. This shows that the optimization of the electrode microstructure is very sensitive to manufacturing parameters and should be done on a multi-criteria basis.


\section{Conclusion}

In this study, we employ the finite element method (FEM) with the software Abaqus to simulate a part of the electrode manufacturing process, comprising particles sedimentation and calendering. While the FEM is restricted to relatively small microstructure volumes due to its expensive numerical cost, it is able to predict interactions and fracturation for particles with arbitrary shapes. This is an attractive benefit of this method compared to the widely spread discrete element method (DEM), that is suited to simulate very large particle assemblies. Here, we simulate calendering with fracturation of an assembly consisting of 100 NMC secondary particles, which presents a good tradeoff between numerical cost and representative volume. Our approach is inspired by methods traditionally used in fracture mechanics to model rock fracturing. By inserting cohesive elements at the interfaces of volumic elements, which represent primary NMC particles, we are able to simulate crack initiation and propagation under high stress conditions. The model does not require prior knowledge of crack locations and avoids the artificial loss of volume associated with element deletion techniques.

The cohesive element parameters are calibrated using experimental nanoindentation data, including both flat-platen and cono-spherical indentation tests, ensuring the model's accuracy at the single-particle level. The model successfully reproduces particle size dependency and demonstrates that the size of primary particles has a minimal impact on fracture properties, such as critical displacement and active material surface.

In larger particle assemblies, our model highlights the mechanism of force chain redistribution, which governs the mechanical properties of the sample during the calendering phase. We observe three distinct regimes: an initial phase dominated by particle rearrangement with minimal fracturing, followed by a phase where particle fracturing occurs under moderate pressure, and finally, a phase where particles are completely crushed, accompanied by a significant increase in stress. 

Despite the computational challenges posed by the insertion of cohesive elements, our model's ability to accurately capture and analyze the post-fracture behavior of secondary NMC particles sets it apart from traditional approaches. In particular, this method captures well micro-structural variables that are essential for electrochemistry. Namely, pressure, porosity, surface, and number of fractured particle are obtained as functions of strain. Their high sensitivity to strain during the intermediate calendering regime opens up perspectives for improving electrode manufacturing through fine-tuning of this process. 

Additionally, the simulation workflow presented here could be extended to anticipate electrochemical performances of an artificial microstructure. The Doyle-Fuller-Newman model (or P2D model) widely used to simulate battery cycling at the electrode scale includes three variables to modulate the effective transport properties of lithium ions in the microstructure, namely the porosity, the tortuosity and the exchange surface. Among them, the porosity and the exchange surface are easily obtained by direct summation of nodes or geometric elements belonging either to the electrolyte or to the solid phase. Only the tortuosity computation requires microstructure meshing and the resolution of a Laplace equation, that could be done in Abaqus at the end of the mechanical simulation. This step will be added in a future work to obtain a complete set of electrochemical parameters for multi-criteria optimization of the microstructure.

\paragraph{Acknowledgements}: We acknowledge support from France 2030 through the project ANR-22-PEBA-0002.


\clearpage

\section{Numerical aspects}

\label{section:methods}

Modeling intra-particle behavior requires a Finite Element software, which distinguishes our work from traditional Discrete Element Methods, in which particles are modeled as plain spheres. Due to its user-friendly interface and wide range of parametric configurations, we chose Abaqus software for microstructure modeling. The computing parts of this software are written in \verb!Fortran! and \verb!C++! for high computing performance, and the interfacing scripts and parametrizations are written in \verb!Python!. Furthermore, parallelism through domain decomposition is fully implemented, which significantly reduces computing time for large-scale simulations.

\subsection{Abaqus basics}

 Since Abaqus has no built-in units, a first step is to choose a complete and consistent set of reference quantities adapted to the physics of the problem. Based on the size of an NMC particle and the usual levels of applied pressure during calendering, we choose one µm and one GPa as base length and pressure units, respectively. Similarly, we choose one fkg ($10^{-15}$ kg) as the base mass unit. From this we deduce the base units for force, density, time and energy, which are respectively one mN, one $\text{fkg.µm}^{-3}$, one ns, and one nJ. Note as an example that in this system of units, the density of water is $1~\text{g.cm}^{-3} = 1~ \text{fkg.µm}^{-3}$.

The interface of Abaqus contains multiple modules allowing the definition of the main problem parameters, based on this system of units.
In the \textbf{part} module  are defined the geometrical caracteristics of the parts involved in the model. In our case, we deal with four different types of parts. First, we have one part for each NMC particle, each of them having a slightly different radius, around $6.5~\text{µm}$. Then, we have two additional parts, one bottom and one upper compression plate. The first one models the current collector, while the other represents the calendering cylinder. Eventually, all the NMC particles are held together laterally with an external mould defined in Abaqus as an analytical rigid surface.  The meshing and selection of the finite element type are also performed in the \textbf{part} module. For computational reasons, we choose an hexaedral mesh associated with the C3D8R element. Indeed, this combination tends to greatly reduce the computational time for a good level of precision, when compared to a tetraedral mesh. Typically, we estimated that the computational time for calendering a single NMC particle could be reduced by a factor of two, for the same level of precision. The \textbf{assembly} module then defines the instantiation of the parts and their relative positions. For example, it specifies the positioning of the compression plates relative to the external mold using geometric constraints.

An other module corresponds to the \textbf{material} definition. Here, we have four materials: the first one is the bulk NMC material (pure elastic, with a Young modulus $E_\mathrm{NMC} = 200~\text{GPa}$ and a Poisson ratio $\nu_\mathrm{NMC} = 0.3$ \cite{stallard2022mechanical} \cite{sharma2023nanoindentation}). The two next materials correspond to the two compression plates. Then, the last material implements the traction-separation law, for the cohesive elements involved in the fracture model presented in the next section. Materials are then assigned to the parts using the \textbf{sections} module. The contact data is defined in the \textbf{interaction} and \textbf{interaction properties} modules. In this model, we adopt the usual Coulomb friction \cite{drucker1954coulomb}, with a hard normal contact behaviour associated to a tangential friction coefficient $\mu$. Finally, the \textbf{loads} and \textbf{BCs} modules allow the definition of the applied loads and boundary conditions.

The analysis type and solver, as well as the related simulation parameters are chosen in the \textbf{steps} module. Although different types of steps are possible, they rely either on an explicit solver (Abaqus explicit) or an implicit solver (Abaqus standard). While the first solver is based on a discretisation of the time domain in a large number of intervals, the second one solves an implicit system of equations using typically a Newton-Raphson method. In Abaqus explicit, the cost per increment is low, but many increments are required, while Abaqus standard uses a small number of expensive steps to converge. Consequently, this makes Abaqus explicit a method of choice for high speed dynamics, and complex contacts, while Abaqus standard is more appropriate for static or quasi static problems. Therefore, we choose here to use Abaqus explicit since our problem involves complex contact and fracture mechanics.
Eventually, the output results of the simulation can be selected in the \textbf{field} and \textbf{history output requests} modules.

\subsection{Fracture models}

This paper focuses on the fracture of an assembly of NMC particules, which first mandates an accurate and reliable fracture model for a  single particle. The Abaqus software offers many options, each of them with their benefits and drawbacks.

A first option is to use progressive damage and failure. This constitutive law is directly implemented in the material properties and is therefore simple to use and has a low computational cost. But since this model mostly applies to ductile materials \cite{eason1996modeling}, it doesn't seem to be relevant for NMC particles.

An other option is to use a brittle cracking model \cite{imseeh20183d}. This works by assigning a brittle material property to the active material definition. Abaqus uses a Rankine criterion for initiating the cracks in the material, meaning that cracks start when the maximum principal stress (i.e, the largest eigenvalue of the stress tensor) is higher than the user-defined tensile strength of the active material.
Then, the crack propagates in the normal to the direction of the maximum tensile principal stress.
Failed elements are removed from the mesh to model the opening of the crack. This model is simple to define and computationally effective, since it can be implemented in the material definition module, but due to the suppression of elements an artificial loss of active material volume is expected, which doesn't meet our requirements. Also, this method imposes mesh dependency on crack propagation.

As an alternative, XFEM (Extended Finite Element Models) \cite{du2009extended} \cite{cruz2019xfem} is a method relying on the addition of discontinuous functions to the usual FE space, for capturing the discontinuous effects of the fracture. As a benefit, crack propagation is not mesh dependent, since it can split between single elements.
As a main drawback, it is mostly applicable to crack propagation and it therefore requires the knowledge of a preexisting crack.

 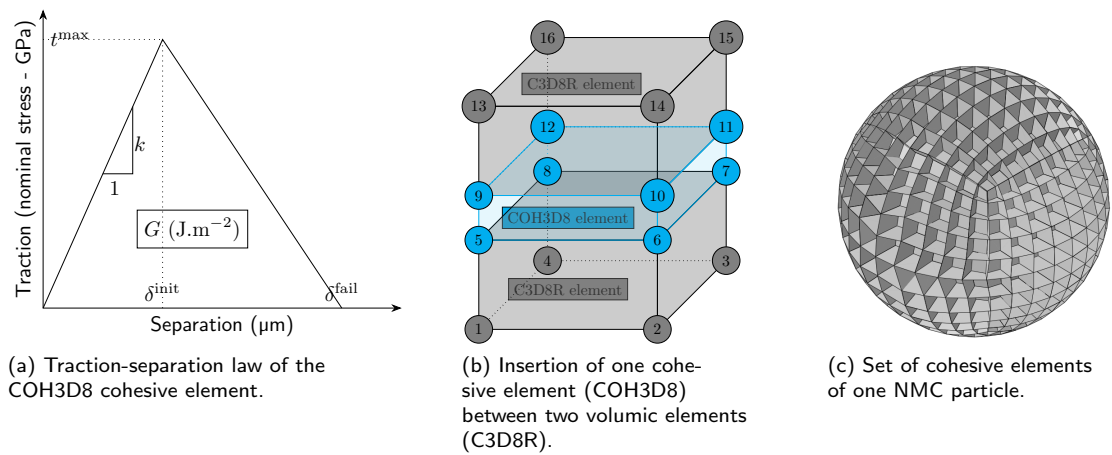
\begin{figure}[ht!]
    \centering
    \subfloat[Traction-separation law of the COH3D8 cohesive element.]{\begin{adjustbox}{clip,trim=0.cm 0.cm 0cm 0cm,max width=0.28\linewidth}\begin{tikzpicture}[
         > = Stealth,
dot/.style = {circle, fill, minimum size=3pt, inner sep=0pt},
  N/.style = {font=\sffamily, align=left},
every label/.style = {label distance = 0pt, N}
                        ]
\draw[<->]  (0,5) |-  (6,0)
    node[pos=0.25, N, above,rotate=90]     {Traction (nominal stress - GPa)}
    node[pos=0.75, N, below]    {Separation (µm)};  
    
    \draw (0,0) -- (2,4.5) -- (5,0);
    \draw[dotted] (0,4.5) -- (2,4.5);
    \draw[dotted] (2,0) -- (2,4.5);
    
    \node[right] at (0, 4.5)   (a) {$t^\mathrm{max}$};
    \node[above] at (2, 0)   (b) {$\delta^\mathrm{init}$};
    \node[above] at (5, 0)   (c) {$\delta^\mathrm{fail}$};
    
    \node[draw, above] at (2.5, 1)   (d) {$G$ ($\text{J.m}^{-2}$)};
    
    \draw (1,4.5/2) -- (1.5,4.5/2) -- (1.5,1.5*4.5/2);
    \node[ below] at (1.2, 4.5/2)   (e) {1};
    \node[right] at (1.4, 4.5/2 + 0.5)   (e) {$k$};

    \end{tikzpicture}
\end{adjustbox}}\quad\quad
     \subfloat[Insertion of one cohesive element (COH3D8) between two volumic elements (C3D8R).]{\begin{adjustbox}{clip,trim=0cm 0.cm 0cm 0cm,max width=0.2\linewidth}\begin{tikzpicture}
\pgfmathsetmacro{\cubex}{4}
\pgfmathsetmacro{\cubey}{2}
\pgfmathsetmacro{\cubez}{4}
\pgfmathsetmacro{\ghost}{1}
\pgfmathsetmacro{\shift}{3}

\node[draw, fill=cyan] at (-2,0.5) {COH3D8 element};
\node[draw, fill=gray] at (-2,-1.1) {C3D8R element};
\node[draw, fill=gray] at (-1.7,3.5) {C3D8R element};

\draw[black,fill=gray, fill opacity=0.4] (0,0,0) -- ++(-\cubex,0,0) -- ++(0,-\cubey,0) -- ++(\cubex,0,0) -- cycle;
\draw[black,fill=gray, fill opacity=0.4] (0,0,0) -- ++(0,0,-\cubez) -- ++(0,-\cubey,0) -- ++(0,0,\cubez) -- cycle;
\draw[black,fill=gray, fill opacity=0.4] (0,0,0) -- ++(-\cubex,0,0) -- ++(0,0,-\cubez) -- ++(\cubex,0,0) -- cycle;

\draw[dotted][black] (-\cubex,-\cubey,0) -- ++(0,0,-\cubez);
\draw[dotted][black](-\cubex,-\cubey,-\cubez) -- ++(0,\cubey,0);
\draw[dotted][black] (-\cubex,-\cubey, -\cubez) -- ++(\cubex,0,0);

\draw[black,fill=gray, fill opacity=0.4] (0,\shift,0) -- ++(-\cubex,0,0) -- ++(0,-\cubey,0) -- ++(\cubex,0,0) -- cycle;
\draw[black,fill=gray, fill opacity=0.4] (0,\shift,0) -- ++(0,0,-\cubez) -- ++(0,-\cubey,0) -- ++(0,0,\cubez) -- cycle;
\draw[black,fill=gray, fill opacity=0.4] (0,\shift,0) -- ++(-\cubex,0,0) -- ++(0,0,-\cubez) -- ++(\cubex,0,0) -- cycle;

\draw[dotted][black] (-\cubex,-\cubey + \shift,0) -- ++(0,0,-\cubez);
\draw[dotted][black](-\cubex,-\cubey + \shift,-\cubez) -- ++(0,\cubey,0);
\draw[dotted][black] (-\cubex,-\cubey + \shift, -\cubez ) -- ++(\cubex,0,0);

\draw[cyan,fill=cyan, fill opacity=0.1] (0, 2*\cubey- \shift,0) -- ++(-\cubex,0,0) -- ++(0,-\shift+\cubey,0) -- ++(\cubex,0,0) -- cycle;
\draw[cyan,fill=cyan, fill opacity=0.1] (0,2*\cubey- \shift,0) -- ++(0,0,-\cubez) -- ++(0,-\shift+\cubey,0) -- ++(0,0,\cubez) -- cycle;
\draw[cyan,fill=cyan, fill opacity=0.1] (0,2*\cubey- \shift,0) -- ++(-\cubex,0,0) -- ++(0,0,-\cubez) -- ++(\cubex,0,0) -- cycle;

\draw[dotted][cyan](-\cubex,2*\cubey- \shift,-\cubez) -- ++(0,\cubey -\shift,0);

\node[state, fill=cyan, minimum size=2pt] at (0,0,0){6};
\node[state, fill=gray, minimum size=2pt] at (0,-\cubey,0){2};
\node[state, fill=gray, minimum size=2pt] at (-\cubex,-\cubey,0){1};
\node[state, fill=gray, minimum size=2pt] at (-\cubex,-\cubey,-\cubez){4};
\node[state, fill=cyan, minimum size=2pt] at (-\cubex,0,-\cubez){8};
\node[state, fill=cyan, minimum size=2pt] at (0,0,-\cubez){7};
\node[state, fill=cyan, minimum size=2pt] at (-\cubex,0,0){5};
\node[state, fill=gray, minimum size=2pt] at (0,-\cubey,-\cubez){3};

\node[state, fill=gray, minimum size=2pt] at (0,\shift,0){14};
\node[state, fill=cyan, minimum size=2pt] at (0,\shift-\cubey,0){10};
\node[state, fill=cyan, minimum size=2pt] at (-\cubex,\shift-\cubey,0){9};
\node[state, fill=cyan, minimum size=2pt] at (-\cubex,\shift-\cubey,-\cubez){12};
\node[state, fill=gray, minimum size=2pt] at (-\cubex,\shift,-\cubez){16};
\node[state, fill=gray, minimum size=2pt] at (0,\shift,-\cubez){15};
\node[state, fill=gray, minimum size=2pt] at (-\cubex,\shift,0){13};
\node[state, fill=cyan, minimum size=2pt] at (0,\shift-\cubey,-\cubez){11};

\end{tikzpicture}
\end{adjustbox}}\quad \quad\quad
\subfloat[Set of cohesive elements of one NMC particle.]{\begin{adjustbox}{clip,trim=0cm 0.cm 0cm 0cm,max width=0.2\linewidth}\includegraphics{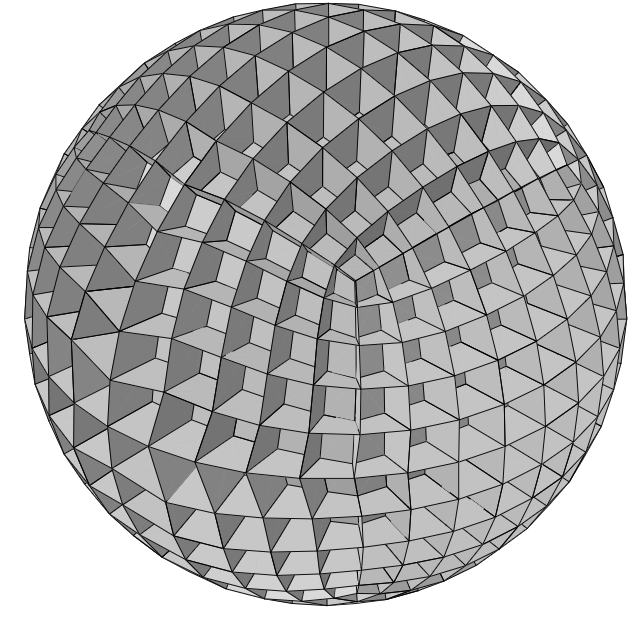} 
\end{adjustbox}}
 \caption{\textbf{The fracture model of a NMC particle:} a cohesive element obeying a traction-separation law (left panel) is inserted at every interface between volumic elements (central panel). A macroscopic view of the resulting set of cohesive elements on a NMC particle is given in the right panel.}
 \label{coh_insert}
\end{figure}

To overcome these drawbacks, we use an other method initially designed for modelling the fracture of rocks \cite{wu2018investigation}. This method works by inserting cohesive elements at the interface between every volumic element of the model. A cohesive element works as a double-sided tape: below a given level of stress, it holds together the two volumic elements it is attached to. However, when stress is higher than the predefined failure level, the cohesive element breaks, meaning that the two previously attached volumic elements separate. One advantage of the method is that it doesn't require knowledge of any preexisting cracks in the material. Also, starting from zero-thickness cohesive elements ensures that no active material volume is lost during the simulation, when the cohesive elements fail. Finally, this choice accurately represents the physics of an NMC particle since a secondary particle is an aggregation of primary particles with a typical size of $\ell =0.5~\text{µm}$ \cite{anansuksawat2024uniform}. Therefore, in this approach, every volumic element models a single primary particle, aggregated into a secondary particle with cohesive elements. As a drawback, our approach requires an important number of cohesive elements, meaning that the computational cost is high. Additionally, it demands the insertion of cohesive elements, which is a more complicated process than simply selecting a material property, as is done in other fracture models.

Figure \ref{coh_insert} shows how cohesive elements are used in our approach. The left panels gives the constitutive traction-separation law of the cohesive elements. As mentioned before, they are inserted at the interface of every volumic element. It therefore requires the duplication of the nodes defining the interface, plotted in blue on the central panel of the figure. Note that the nodes are just duplicated and not displaced, meaning that the cohesive element has a zero-thickness. Technically speaking, this insertion is achieved by running a python code in the Abaqus interface. This code gets as an input the mesh results of a previous Abaqus simulation, in other word the position of the nodes and the elements connectivity. Then, it computes all the couples of volumic elements that share a common interface. The complexity of this operation, which is quadratic in the number of elements, is lowered by the use of a hash table ordering all the faces in accordance with a user-defined key. Once this data is gathered, the new mesh is constructed from scratch and integrates the duplicated interface nodes. The right panel of figure \ref{coh_insert} gives the macroscopic result of this process, showing the set of all cohesive elements of a NMC particle. This is a visual validation of the insertion process presented above.

\bibliographystyle{MSP}
\bibliography{biblio.bib}

\end{document}